\documentclass[english,superscriptaddress,twocolumn]{revtex4-2}
\usepackage[T1]{fontenc}
\usepackage[latin9]{inputenc}
\usepackage{amsmath}
\usepackage{amssymb}
\usepackage{cancel}
\usepackage{graphicx}
\usepackage{babel}
\usepackage{xcolor}
\usepackage{slashed}
\usepackage{comment}

\begin{document}

\title{Spectral Functions of $J/\psi$ Meson in Rotating Thermal Background from Holography}

\author{Xin-Li Sheng}
\affiliation{Shanghai Research Center for Theoretical Nuclear Physics, 
NSFC and Fudan University, Shanghai 200438, China}

\author{Jun-Xia Chen}
\affiliation{College of Physics and Electronic Science,
	Hubei Normal University,
	Huangshi 435002, China}

\author{Defu Hou}
\affiliation{Institute of Particle Physics and Key Laboratory of Quark and Lepton Physics (MOS),
	Central China Normal University,
  Wuhan 430079, China}

\author{Hai-Cang Ren}
\affiliation{School of Physical Science and Technology,  ShanghaiTech University,
Shanghai 201210, China}

\begin{abstract}
    We investigate the spectral functions of the $J/\psi$ meson in a rotating thermal background within the soft-wall holographic model. The global rotation is implemented through a rotating AdS-like metric, while a local inertial frame is introduced in which the vector field can be decomposed into different spin states. We solve the equations of motion of the vector field in the bulk with incoming wave condition near the horizon and compute the retarded Green function, from which we extract the invariant-mass spectral functions for $J/\psi$. When the momentum is parallel to the rotation axis, the peak energies shift by $-\Omega J_z$, as expected by a coupling between rotation and angular momentum, while the widths are nearly independent to $\Omega$. When the momentum is in perpendicular direction, the spectral functions for spin-$\pm1$ states deviate significantly from the single-peak behavior and the energy shifts depart from $-\Omega J_z$. The resulting triplet splittings of the spectral functions provides a holographic perspective on spin-dependent vector meson properties in rotating systems.
\end{abstract}

\maketitle

\section{Introduction}

Relativistic heavy-ion collisions provide a unique platform for studying strongly interacting matter under extreme conditions \cite{Rischke:2003mt,Gyulassy:2004zy,Shuryak:2004cy}. In non-central collisions, the two nuclei carry a sufficiently large initial orbital angular momentum, part of which has been transferred to the produced quark-gluon plasma (QGP), leading to significant global rotation perpendicular to the event plane. This vortical environment gives rise to various spin-related phenomena through the particle's spin-orbit coupling, such as the global spin polarization of hyperons \cite{Voloshin:2004ha,Liang:2004xn,Liang:2004ph}, the chiral vortical effect \cite{Banerjee:2008th,Torabian:2009qk,Son:2009tf,Kharzeev:2015znc,Huang:2017pqe}, the chiral vortical wave \cite{Jiang:2015cva,Kharzeev:2015znc,Huang:2017pqe}, etc. In experiments, the global polarization of $\Lambda$ has been observed by the STAR collaboration \cite{STAR:2007ccu,STAR:2017ckg,STAR:2021beb}, providing direct evidence for the vorticity field in the QGP, and can be reproduced by many model simulations \cite{Karpenko:2016jyx,Li:2017slc,Guo:2021udq,Sun:2017xhx,Becattini:2024uha}. On the other hand, it is proposed that the global quark polarization induced by the vorticity field could contribute to vector meson spin alignment \cite{Liang:2004xn,STAR:2022fan}. The non-relativistic coalescence model predicts that the meson's spin density matrix element $\rho_{00}$ will deviate from $1/3$ by $-(1/9)\Omega^2/T^2$ \cite{Yang:2017sdk}, where $\Omega$ denotes the angular velocity. Although such a contribution is expected to be small and may not be the dominant mechanism for the vector meson spin alignment \cite{Sheng:2019kmk,Sheng:2022wsy,Sheng:2023urn,Chen:2024afy}, understanding how the rotation modifies meson properties in a strongly-coupled system is still an important step towards establishing a comprehensive description of vector mesons in heavy-ion collisions. 

For the QCD matter in the strongly-coupled regime, the perturbative approach becomes unreliable. The AdS/CFT correspondence \cite{Maldacena:1997re, Witten:1998zw,Witten:1998qj,Aharony:1999ti}, on the other hand, serves as a powerful method in this regime, with various holographic models being developed for the QCD matter, e.g., top-down approaches based on D-brane \cite{Karch:2002sh, Sakai:2004cn,Sakai:2005yt} and bottom-up approaches such as the soft wall model and hard wall model \cite{Karch:2006pv,Gherghetta:2009ac,Fujita:2009wc, Mamani:2013ssa,Braga:2017bml,Erlich:2005qh,Herzog:2006ra}. Within these models, physical quantities are mapped to bulk fields in the dual gravity theory in a higher-dimensional curved spacetime, which are computed perturbatively, such as confinement \cite{Andreev:2009zk,  Andreev:2006nw, Braga:2022yfe, Chen:2020ath, Wang:2024szr, Zhao:2022uxc, Chen:2024edy}, potential \cite{Rey:1998bq, Brandhuber:1998bs, Rougemont:2014efa, Zhang:2016fdk, Chen:2022obe}, energy loss \cite{Liu:2006ug, Herzog:2006gh, Hou:2021own, Chen:2023yug, Zhu:2021nbl}, and so on. Remarkably, heavy quarkonium properties, described by the spectral functions, are studied through the two-point retarded Green functions of the dual current \cite{Son:2002sd, Policastro:2002se,kim2007heavy,martin2021heavy, Fujita:2009wc,fujita2010melting, Sheng:2024kgg, Zhao:2024ipr}. Recent holographic studies have further extended these models to include rotation, revealing how rotation influences the spectral function of heavy quarkonium \cite{Braga:2023fac, Zhao:2023pne, Zhu:2024uwu,Wang:2024rim}. 

In this work, we focus on the spectral properties of the heavy quarkonium $J/\psi$ meson in a rotating system within the QCD soft-wall model \cite{Karch:2006pv,Gherghetta:2009ac,Fujita:2009wc,Mamani:2013ssa,Braga:2017bml}. The equations of motion for the bulk vector field are derived in the local inertial frame using local vielbein, which enables a transparent description of the coupling between spin and rotation. Near the horizon, the bulk vector field is solved analytically up to the next-to-leading order in the distance to the horizon. The equations of motion are then solved numerically in the whole bulk region, allowing us to calculate the retarded Green function. The spectral functions for spin-$0,\pm 1$ mesons are extracted from the Green function by first projecting it onto polarization vectors and then taking the imaginary part. The obtained spectral functions, especially the peak masses, exhibit clear dependence on the global rotation, which arises from the coupling between the angular momentum of the vector meson and the rotational background. These results provide a holographic perspective on the spin-dependent vector meson properties in a vortical strongly-coupled QCD matter and may offer useful insights on probing rotations with heavy quarkonium. 

The paper is organized as follows. In Sec. \ref{sec:EOM}, we introduce the holographic setup and derive the equations of motion for the vector field in the rotating background. The solutions near the horizon are constructed in Sec. \ref{sec:Near-Horizon}. The Green function and spectral functions are expressed in terms of the vector field in Sec. \ref{sec:Spectral-Function}, with numerical results being presented in Sec. \ref{sec:Numerical-results}. Finally, a summary is given in Sec. \ref{sec:summary}. 

\section{Equations of motion}\label{sec:EOM}

According to the AdS/CFT correspondence, a vector meson in a strongly coupled system can be described by the boundary value of a bulk $U(1)$ gauge field $A_{M}(x,\zeta)$ in the AdS-like space. We choose the corresponding bulk mesonic action as the generalized Maxwell form
\begin{equation}\label{eq:action-vector-field}
S_\text{bulk}=-\frac{1}{4g_{5}^{2}}\int d^{4}x\,d\zeta\,\sqrt{-g}\,e^{-\Phi(\zeta)}F_{MN}F^{MN}\,,
\end{equation}
where $F_{MN}\equiv\partial_M A_N-\partial_N A_M$ is the field strength tensor and $M,\,N$ run over the five-dimensional bulk. The Yang-Mills coupling constant $g_5^2=12\pi^2L/N_c$ is determined by matching the UV asymptotic behavior of the current correlator to the dual QFT \cite{Erlich:2005qh}. Here $\Phi(\zeta)$ is the dilaton field which plays the role of infrared cut-off and will reproduce vector meson masses. In this work we take
\begin{equation}
\Phi(\zeta)=c\zeta^2
\end{equation}
according to the soft-wall model \citep{Karch:2006pv,Mamani:2013ssa}, with the parameter $c=m_{J/\psi}^2/4\approx2.40\,\text{GeV}^2$. Here and in the follows, we restrict the type of vector meson as the charmonium $J/\psi$, while the whole framework can be easily extended to $\phi$ or bottomonium by changing the parameter $c$ in the dilaton field.  

For simplicity, we define 
\begin{equation}
Q(\zeta)\equiv-\frac{1}{g_5^2}\sqrt{-g}e^{-\Phi(\zeta)}\,,
\end{equation}
and the equation of motion for the bulk field is given by 
\begin{equation}\label{equation-of-motion}
\partial_M\left[g^{MJ}g^{NK}Q(\zeta)F_{JK}\right]=0\,.
\end{equation}
In the absence of rotation, the background geometry is taken to be asymptotically AdS, with the spacetime interval
\begin{equation}
ds^{2}=\frac{L^{2}}{\zeta^{2}}\left[-f(\zeta)dt^{2}+dx^{2}+dy^{2}+dz^{2}+\frac{d\zeta^{2}}{f(\zeta)}\right]\,,
\end{equation}
where $L$ denotes the AdS radius and $f(\zeta)\equiv1-(\zeta/\zeta_h)^4$ with $\zeta_h$ being the location of horizon. A global rotation with angular velocity $\Omega$ drifts the azimuthal angle as $\phi\rightarrow\phi+\Omega t$ and thus the spacetime interval becomes 
\begin{eqnarray}
ds^{2}&=&\frac{L^{2}}{\zeta^{2}}\Big[-f(\zeta)dt^{2}+dx^{2}+dy^{2}+dz^{2}+\frac{d\zeta^{2}}{f(\zeta)} \nonumber \\
&&+2\Omega(xdy-ydx)dt+\Omega^{2}(x^{2}+y^{2})dt^{2}\Big]\,,
\end{eqnarray}
corresponding to the following metric
\begin{align}\label{eq:metric-with-rotation}
&g_{MN}(x)\nonumber\\
&=\frac{L^{2}}{\zeta^{2}}\left(\begin{array}{ccccc}
-f(\zeta)+\Omega^{2}(x^{2}+y^{2}) & -\Omega y & \Omega x & 0 & 0\\
-\Omega y & 1 & 0 & 0 & 0\\
\Omega x & 0 & 1 & 0 & 0\\
0 & 0 & 0 & 1 & 0\\
0 & 0 & 0 & 0 & 1/f(\zeta)
\end{array}\right).
\end{align}
The Hawking temperature is determined by \citep{zhao:1983hawking}
\begin{equation}
T=\frac{|\kappa|}{2\pi}=\frac{1}{\pi\zeta_h}\,,
\end{equation}
where $\kappa$ denotes the surface gravity. 

We note that the rotational effect is included in the equation of motion (\ref{equation-of-motion}) through the metric (\ref{eq:metric-with-rotation}). For the vector field considered in this paper, which has spin-1, Eq.~(\ref{equation-of-motion}) is incomplete at a conceptual level. This is because spin is defined as a representation of the local Lorentz transformation, rather than the general coordinate transformation encoded by the metric. At the AdS boundary $\zeta=0$, the metric (\ref{eq:metric-with-rotation}) does not reduce to the Minkowski one $\eta_{\mu\nu}=\text{diag}(-1,1,1,1)$, indicating that the spin of the bulk field is not a well-defined quantity at the boundary. In order to more explicitly describe spin and its coupling with rotation, it is necessary to introduce the vielbein $e_{\ \ M}^{a}(x)$ which relates the curved metric to a locally flat Minkowski metric,
\begin{equation}
g_{MN}(x,\zeta)=\eta_{ab}(\zeta)e_{\ M}^{a}(x)e_{\ N}^{b}(x)\,.
\end{equation}
Here $\eta_{ab}=(L^{2}/\zeta^{2})\text{diag}(-f(\zeta),1,1,1,1/f(\zeta))$ denotes the standard AdS metric which has a flat 4-d boundary. The nonzero components of the locally-defined vielbein are taken as follows
\begin{equation}
e_{\ t}^{t}=e_{\ x}^{x}=e_{\ y}^{y}=e_{\ z}^{z}=e_{\ \zeta}^{\zeta}=1,\ e_{\ t}^{x}=-\Omega y,\ e_{\ t}^{y}=\Omega x\,,
\end{equation}
and the vector field in the rotating background is related to that in the local inertial frame as
\begin{equation}
A_{M}(x,\zeta)=e_{\ M}^{a}(x)A_{a}^{\text{local}}(x,\zeta)\,.
\end{equation}
It is obvious that $A_a^\text{local}$ coincides with $A_M$ in the absence of rotation. We then impose the radial gauge condition $A_\zeta^\text{local}=0$ and take the Fourier transformation for the remaining components,
\begin{equation}
A_{\mu}^{\text{local}}(x,\zeta)=\int\frac{d\omega d^{3}{\bf q}}{(2\pi)^{4}}e^{-i\omega t+i{\bf q}\cdot{\bf x}}\tilde{A}_{\mu}(q,\zeta)\,.
\end{equation}
The vector field $A_{M}(x,\zeta)$ is then expressed as 
\begin{align}\label{eq:Fourier-A_M}
A_{M}(x,\zeta)=&\int\frac{d\omega d^{3}{\bf q}}{(2\pi)^{4}}e^{-i\omega t+i{\bf q}\cdot{\bf x}}\nonumber\\
&\hspace{-1cm}\times\left(\tilde{A}_{t}+\Omega(x\tilde{A}_{y}-y\tilde{A}_{x}),\,\tilde{A}_{x},\,\tilde{A}_{y},\,\tilde{A}_{z},\,0\right)\,,
\end{align}
where $\tilde{A}_\mu$, $\mu=t,x,y,z$ are functions of $\omega$, ${\bf q}$, and $\zeta$. Due to the global rotation, the metric $g^{MN}$ also explicitly depends on the transverse coordinates $x$ and $y$. This indicates that when substituting Eq. (\ref{eq:Fourier-A_M}) into the equation of motion (\ref{equation-of-motion}), we will face unexpected mixing between different Fourier modes, making the system nearly impossible to solve. To avoid this difficulty, we assume that a particle is treated as a plane wave at microscopic scales but as a point particle at macroscopic scales. Such an assumption would be valid if the scale of the inhomogeneity of metric is much larger than the particle's wavelength, i.e.,
\begin{equation}\label{eq:approximation}
\frac{1}{\Omega}\gg\frac{1}{\omega},\,\frac{1}{|\bf{q}|}\,,
\end{equation}
or equivalently,
\begin{equation}
\omega,\,|{\bf q}|\gg \Omega\,.
\end{equation}
In real heavy-ion collisions, the angular velocity of global rotation is just a few MeV, while the momentum of the considered particles is of the order of GeV. In this paper, we consider $\Omega$ up to $0.1$ GeV, which is still much smaller than the $J/\psi$ mass. Therefore the above assumption is reasonable in both heavy-ion collisions and in this paper, allowing us to choice a classical transverse location $(x,y)$ for the particle and the spacetime geometry felt by the particle is determined by the metric at $(x,y)$.   
We are then able to derive the equations of motion for different Fourier modes
\begin{align}\label{eq:EOM}
0	&=\ 	Qg^{\zeta\zeta}g^{\alpha\beta}\left[\tilde{A}_{\beta}^{\prime\prime}+\Omega\delta_{\beta t}(x\tilde{A}_{y}^{\prime\prime}-y\tilde{A}_{x}^{\prime\prime})\right] \nonumber\\
		&+(\partial_{\zeta}Qg^{\zeta\zeta}g^{\alpha\beta})\left[\tilde{A}_{\beta}^{\prime}+\Omega\delta_{\beta t}(x\tilde{A}_{y}^{\prime}-y\tilde{A}_{x}^{\prime})\right]\nonumber\\
		&-Qg^{\mu\nu}g^{\alpha\beta}q_{\mu}(q_{\nu}\tilde{A}_{\beta}-q_{\beta}\tilde{A}_{\nu})\nonumber\\
		&-\Omega Qg^{\mu\nu}g^{\alpha\beta}q_{\mu}(q_{\nu}\delta_{\beta t}-q_{\beta}\delta_{\nu t})(x\tilde{A}_{y}-y\tilde{A}_{x})\nonumber\\
		&+i\Omega Qg^{\mu\nu}g^{\alpha\beta}\left[(q_{\nu}\delta_{\beta t}-q_{\beta}\delta_{\nu t})(\delta_{\mu x}\tilde{A}_{y}-\delta_{\mu y}\tilde{A}_{x})\right.\nonumber\\
		&\left.-q_{\mu}(\delta_{\nu y}\delta_{\beta t}-\delta_{\beta y}\delta_{\nu t})\tilde{A}_{x}+q_{\mu}(\delta_{\nu x}\delta_{\beta t}-\delta_{\beta x}\delta_{\nu t})\tilde{A}_{y}\right]\nonumber\\
		&+iQ(\partial_{\mu}g^{\mu\nu}g^{\alpha\beta})(q_{\nu}\tilde{A}_{\beta}-q_{\beta}\tilde{A}_{\nu})\nonumber\\
		&+i\Omega Q(\partial_{\mu}g^{\mu\nu}g^{\alpha\beta})(q_{\nu}\delta_{\beta t}-q_{\beta}\delta_{\nu t})(x\tilde{A}_{y}-y\tilde{A}_{x})\nonumber\\
		&+\Omega Q(\partial_{\mu}g^{\mu\nu}g^{\alpha\beta})\nonumber\\
        &\times\left[(\delta_{\nu x}\delta_{\beta t}-\delta_{\beta x}\delta_{\nu t})\tilde{A}_{y}-(\delta_{\nu y}\delta_{\beta t}-\delta_{\beta y}\delta_{\nu t})\tilde{A}_{x}\right]\,,
\end{align}
and a constraint equation for the first-order derivatives 
\begin{align}\label{eq:constraint}
0\hspace{-0.1cm}=&Qg^{\zeta\zeta}\left\{ g^{\mu\nu}q_{\mu}\tilde{A}_{\nu}^{\prime}-i(\partial_{\mu}g^{\mu\nu})\left[\tilde{A}_{\nu}^{\prime}+\Omega\delta_{\nu t}(x\tilde{A}_{y}^{\prime}-y\tilde{A}_{x}^{\prime})\right]\right.\nonumber\\
&\left.-i\Omega g^{\mu\nu}\delta_{\nu t}\left[iq_{\mu}(x\tilde{A}_{y}^{\prime}-y\tilde{A}_{x}^{\prime})+(\delta_{\mu x}\tilde{A}_{y}^{\prime}-\delta_{\mu y}\tilde{A}_{x}^{\prime})\right]\right\}\,.
\end{align}
In these equations, $g^{\mu\nu}$ is a function of coordinates $x$, $y$, and $\zeta$, and $\tilde{A}_\mu$ is a function of $q_\mu=(-\omega,\,{\bf q})$. We have embedded the explicit parameter dependence for simplicity and labeled $\tilde{A}_{\mu}^{\prime}\equiv\partial_{\zeta}\tilde{A}_{\mu}$, $\tilde{A}_{\mu}^{\prime\prime}\equiv\partial_{\zeta}^{2}\tilde{A}_{\mu}$. Eqs. (\ref{eq:EOM}) and (\ref{eq:constraint}) reduce to those in Refs. \citep{Mamani:2013ssa,Sheng:2024kgg} when $\Omega=0$. In this paper, we quantify the bulk behavior of a quasi-classical vector meson with momentum $q_\mu$ and located at $(x,\,y)$ on the transverse plane. We will show later that such a meson has an energy shift of $-({\bf x}\times{\bf q}+\bf s)\cdot\boldsymbol{\Omega}$ due to the coupling between its total angular momentum and the global rotation, as predicted by classical theories.  

\section{Solution near horizon}\label{sec:Near-Horizon}
Similar to the method in the absence of global rotation \citep{Mamani:2013ssa,Sheng:2024kgg}, we define the electric fields in the momentum space as linear combinations of $\tilde{A}_\mu(q,\zeta)$,
\begin{equation}\label{eq:def-E}
E_{i}(q,\zeta)\equiv\omega\tilde{A}_{i}(q,\zeta)+q_{i}\tilde{A}_{t}(q,\zeta)\,,\ \ \ i=x,\,y,\,z\,.
\end{equation}
The convention of $E_i$ differs from the traditional electric field in classical electrodynamics by a factor of $i$. Since here $E_i$ just serves as an auxiliary field for simplifying our calculation, this difference will not affect our result of the meson's spectral function and therefore we still call $E_i$ the electric field. We introduce column vectors $\mathcal{E}\equiv(E_{t},E_{x},E_{y},E_{z})^{T}$ and $\mathcal{A}\equiv(\tilde{A}_{t},\tilde{A}_{x},\tilde{A}_{y},\tilde{A}_{z})^{T}$, where $E_{t}\equiv \omega\tilde{A}_{t}$, such that the relations between $\tilde{A}_i$ and $E_i$ can be expressed in a matrix form
\begin{equation}\label{eq:relation-E-A}
\mathcal{E}=\mathcal{P}_{\mathcal{EA}}\mathcal{A},\ \ \mathcal{A}=\mathcal{P}_{\mathcal{EA}}^{-1}\mathcal{E}\,,
\end{equation}
where
\begin{equation}
\mathcal{P}_{\mathcal{EA}}\equiv\left(\begin{array}{cccc}
\omega & 0 & 0 & 0\\
q_{x} & \omega & 0 & 0\\
q_{y} & 0 & \omega & 0\\
q_{z} & 0 & 0 & \omega
\end{array}\right)\,.
\end{equation}
Using the column vector, Eq. (\ref{eq:EOM}) is reorganized in a matrix form,
\begin{equation}\label{eq:EOM-A}
\mathcal{T}_{1}\mathcal{A}^{\prime\prime}(q,\zeta)+\mathcal{T}_{2}\mathcal{A}^{\prime}(q,\zeta)+\mathcal{T}_{3}\mathcal{A}(q,\zeta)=0\,,
\end{equation}
where $\mathcal{T}_{1,2,3}$ are $4\times4$ matrices that depend on $x$, $y$, $\zeta$, and $q_\mu$. The explicit expressions of $\mathcal{T}_{1,2,3}$ can be extracted from Eq. (\ref{eq:EOM}) and are shown in the Appendix \ref{Appendix}. Using Eq. (\ref{eq:relation-E-A}), the above equation can be converted to a matrix-form equation for $\mathcal{E}$, 
\begin{equation}\label{eq:EOM-E}
\mathcal{E}^{\prime\prime}(q,\zeta)+\mathcal{F}\mathcal{E}^{\prime}(q,\zeta)+\mathcal{G}\mathcal{E}(q,\zeta)=0\,,
\end{equation}
where $\mathcal{F}$ and $\mathcal{G}$ are $4\times4$ matrices given by 
\begin{equation}
\mathcal{F}\equiv\mathcal{P}_{\mathcal{EA}}\mathcal{T}_{1}^{-1}\mathcal{T}_{2}\mathcal{P}_{\mathcal{EA}}^{-1},\ \ \ \mathcal{G}\equiv\mathcal{P}_{\mathcal{EA}}\mathcal{T}_{1}^{-1}\mathcal{T}_{3}\mathcal{P}_{\mathcal{EA}}^{-1}\,.
\end{equation}
On the other hand, Eq. (\ref{eq:constraint}) provides another constraint for $\mathcal{E}^\prime$, 
\begin{equation}\label{eq:constraint-E}
\mathcal{D}\mathcal{E}^{\prime}(q,\zeta)=0\,,
\end{equation}
where $\mathcal{D}$ is a row vector with four-elements. Note that $\mathcal{F}$, $\mathcal{G}$, and $\mathcal{D}$ are functions of $x$, $y$, $\zeta$, and $q_\mu$, and their parameter dependence is embedded in the above equations for simplicity.

In order to analyze the spectral properties, we need to figure out the incoming wave solution for the equation of motion (\ref{eq:EOM-E}) with the constraint (\ref{eq:constraint-E}). We assume that such a solution takes the following form,
\begin{equation}\label{eq:incoming-wave-solution}
\mathcal{E}(q,\zeta)=e^{-i\tilde{\omega}r_{\ast}}\psi(q,\zeta)\,,
\end{equation}
where the tortoise coordinate is defined 
\begin{equation}
r_{*}=\frac{\zeta_{h}}{2}\left[-\arctan\frac{\zeta}{\zeta_{h}}+\frac{1}{2}\ln\frac{\zeta_{h}-\zeta}{\zeta_{h}+\zeta}\right]\,.
\end{equation}
The column vector $\psi(q,\zeta)$ is assumed to be finite at the horizon, while the divergence near the horizon is controlled by $\exp(-i\tilde\omega r_\ast)$ with $\tilde{\omega}>0$. A solution with near-horizon behavior $\sim \exp(i\tilde{\omega}r_\ast)$ is identified as an outgoing solution, which is neglected in this paper because it does not contribute to the meson's spectral function. Substituting (\ref{eq:incoming-wave-solution}) into Eq. (\ref{eq:EOM-E}), we derive the equation for $\psi(q,\zeta)$,
\begin{equation}\label{eq:EOM-psi}
\psi^{\prime\prime}(q,\zeta)+\tilde{\mathcal{F}}\psi^{\prime}(q,\zeta)+\tilde{\mathcal{G}}\psi(q,\zeta)=0\,,
\end{equation}
where
\begin{equation}
\tilde{\mathcal{F}}\equiv\mathcal{F}+\frac{2i\tilde{\omega}}{f(\zeta)},\ \  \tilde{\mathcal{G}}\equiv\mathcal{G}+
\frac{i\tilde{\omega}}{f(\zeta)}\tilde{\mathcal{F}}+\frac{\tilde{\omega}^{2}-i\tilde{\omega}\partial_{\zeta}f(\zeta)}{[f(\zeta)]^{2}}\,.
\end{equation}
Meanwhile, the constraint equation (\ref{eq:constraint-E}) requires that 
\begin{equation}\label{eq:constraint-psi}
\mathcal{D}\left[\psi^{\prime}(q,\zeta)+\frac{i\tilde{\omega}}{f(\zeta)}\psi(q,\zeta)\right]=0\,.
\end{equation}
In a small region near the horizon, the solution can be expressed as a Taylor expansion series
\begin{equation}\label{eq:expansion-psi}
\psi(q,\zeta)=\psi^{(0)}(q)+\sum_{n=1}^{\infty}\left(\frac{\zeta}{\zeta_h}-1\right)^{n}\psi^{(n)}(q)\,,
\end{equation}
where $\psi^{(n)}$ are column vectors and $\psi^{(0)}$ is the value of $\psi(q,\zeta)$ at the horizon $\zeta=\zeta_h$. Similarly, we expand $\tilde{\mathcal{F}}$, $\tilde{\mathcal{G}}$, and $\mathcal{D}$ as
\begin{align}\label{eq:expansion-F-G-D}
\tilde{\mathcal{F}}=&\sum_{n=-1}^\infty\left(\frac{\zeta}{\zeta_h}-1\right)^n\,\tilde{\mathcal{F}}^{(n)},\nonumber\\
\tilde{\mathcal{G}}=&\sum_{n=-2}^\infty\left(\frac{\zeta}{\zeta_h}-1\right)^n\,\tilde{\mathcal{G}}^{(n)},\nonumber\\
\mathcal{D}=&\sum_{n=0}^\infty\left(\frac{\zeta}{\zeta_h}-1\right)^n\,\mathcal{D}^{(n)}\,,
\end{align}
where the lowest orders in the expansion are determined by asymptotic behaviors of these functions near the horizon. Substituting Eqs. (\ref{eq:expansion-psi}) and (\ref{eq:expansion-F-G-D}) into Eqs. (\ref{eq:EOM-psi}) and (\ref{eq:constraint-psi}), we obtain the following equations at lowest order in $\zeta/\zeta_h-1$,
\begin{equation}\label{eq:psi-0}
\tilde{\mathcal{G}}^{(-2)}\psi^{(0)}=0,\ \ \mathcal{D}^{(0)}\psi^{(0)}=0\,,
\end{equation}
where 
\begin{align}\label{eq:coefficients-G2-D0}
\tilde{\mathcal{G}}^{(-2)}=&\frac{1}{16}\left(\begin{array}{cccc}
-\frac{\tilde{\omega}(\tilde{\omega}\zeta_{h}-4i)}{\zeta_{h}} & 0 & 0 & 0\\
\frac{4iq_{x}\tilde{\omega}}{\omega\zeta_{h}} & \omega^{2}-\tilde{\omega}^{2} & 0 & 0\\
\frac{4iq_{y}\tilde{\omega}}{\omega\zeta_{h}} & 0 & \omega^{2}-\tilde{\omega}^{2} & 0\\
\frac{4iq_{z}\tilde{\omega}}{\omega\zeta_{h}} & 0 & 0 & \omega^{2}-\tilde{\omega}^{2}
\end{array}\right)\nonumber \\
&+\frac{\Omega}{16}\left(\begin{array}{cccc}
0 & 0 & 0 & 0\\
iq_{y}-q_{x}L_{z} & 2\omega L_{z} & -2i\omega & 0\\
-iq_{x}-q_{y}L_{z} & 2i\omega & 2\omega L_{z} & 0\\
-q_{z}L_{z} & 0 & 0 & 2\omega L_{z}
\end{array}\right)\,, \nonumber\\
\mathcal{D}^{(0)}=&\frac{\zeta_h^4 Q(\zeta_h) (\omega+\Omega L_z)}{\omega L^4} (1,\,0,\,0,\,0)\,.
\end{align}
Here $L_z\equiv xq_y-yq_x$ denotes the orbital angular momentum in the $z$-direction. It is obvious that Eq. (\ref{eq:psi-0}) has three linear independent solutions,
\begin{equation}
\psi_{0}^{(0)}=\left(\begin{array}{c}
0\\ 0\\ 0\\ 1
\end{array}\right),\ \psi_{1}^{(0)}=\frac{1}{\sqrt{2}}\left(\begin{array}{c}
0\\ 1\\ i\\ 0
\end{array}\right),\ \psi_{-1}^{(0)}=\frac{1}{\sqrt{2}}\left(\begin{array}{c}
0\\ 1\\ -i\\ 0
\end{array}\right)\,.
\end{equation}
They coincide with the polarization vectors for the linearly polarized state and two circularly polarized states, respectively, with the subscript $\lambda=0,\,\pm1$ denotes the spin projection in the $z$-direction. The value of  $\tilde\omega$ corresponding to the solution $\psi^{(0)}_\lambda$ reads
\begin{equation}
\tilde{\omega}_\lambda\simeq\omega+(L_z+\lambda)\Omega+\mathcal{O}(\Omega^2)\,,
\end{equation}
where the difference between $\omega$ and $\tilde\omega$ is exactly the coupling between total angular momentum and the global rotation. 

At next-to-leading order in $\zeta/\zeta_h-1$, Eq. (\ref{eq:EOM-psi}) gives
\begin{equation}
\left[\tilde{\mathcal{G}}^{(-2)}+\tilde{\mathcal{F}}^{(-1)}\right]\psi^{(1)}+\tilde{\mathcal{G}}^{(-1)}\psi^{(0)}=0\,,
\end{equation}
which has a solution
\begin{equation}
\psi^{(1)}=-\left[\tilde{\mathcal{G}}^{(-2)}+\tilde{\mathcal{F}}^{(-1)}\right]^{-1}\tilde{\mathcal{G}}^{(-1)}\psi^{(0)}\,.
\end{equation}
Explicit expressions of $\tilde{\mathcal{F}}^{(-1)}$ and $\tilde{\mathcal{G}}^{(-1)}$ are shown in the Appendix. We then obtain three independent incoming wave solutions, which have the following forms near the horizon,
\begin{align}\label{eq:solution-near-horizon}
\mathcal{E}_\lambda^\text{nh}(q,\zeta)=&\Bigg\{1-\left(\frac{\zeta}{\zeta_h}-1\right)\left[\tilde{\mathcal{G}}^{(-2)}+\tilde{\mathcal{F}}^{(-1)}\right]^{-1}\tilde{\mathcal{G}}^{(-1)}
\nonumber\\
&\hspace{-0.5cm}\left.+\mathcal{O}\left[\left(\frac\zeta\zeta_h-1\right)^2\right]\Bigg\}\psi^{(0)}_\lambda e^{-i\tilde{\omega}r_\ast}\right|_{\tilde{\omega}=\omega+(L_z+\lambda)\Omega}\,.
\end{align}
The high-order corrections in $\zeta/\zeta_h-1$ can be derived order by order in a similar way as what we did for $\psi^{(1)}$. 
In order to calculate the value of $\mathcal{E}_\lambda$ for a $\zeta$ far from the horizon, we need to propagate the incoming wave from the horizon towards the boundary. 
This is achieved by numerically solving the second-order differential equation (\ref{eq:EOM-E}) with initial values $\mathcal{E}(q,\zeta)$ and $\mathcal{E}^\prime(q,\zeta)$ near the horizon. In practice, we take $\zeta/\zeta_h-1\lesssim10^{-6}$ such that high-order corrections in Eq. (\ref{eq:solution-near-horizon}) are significantly suppressed. Equation (\ref{eq:solution-near-horizon}) is then effective for providing high-precision near-horizon values of $\mathcal{E}(q,\zeta)$ and $\mathcal{E}^\prime(q,\zeta)$.

\section{Spectral functions}\label{sec:Spectral-Function}
The spectral function of a vector meson ($J/\psi$ in this paper) is related to the imaginary part of the retarded current correlation function. For the vector field on the boundary, we introduce the current $J^a(x)$ which couples to the vector field in local inertial frame $A_a^\text{local}(x,\zeta=0)$. The retarded current correlation function is evaluated as follows,
\begin{align}
D^{ab}(q)\equiv&-i\int d^4x\,e^{-iq\cdot x}\,\theta(t)\left\langle\left[J^a(x),J^b(0)\right]\right\rangle\nonumber\\
=&-2\frac{\delta^{2}S}{\delta\tilde{A}_{a}(-q,0)\delta\tilde{A}_{b}(q,0)}\,.
\end{align}
Substituting the Fourier expansion (\ref{eq:Fourier-A_M}) into the action in Eq. (\ref{eq:action-vector-field}) and taking second order variation with respect to $\tilde{A}_\mu$, we obtain 
\begin{equation}\label{eq:current-correlation}
D^{ab}(x,y;q)=\frac{1}{(2\pi)^{4}}\left.Q(\zeta)g^{\zeta\zeta}g^{\alpha\beta}e_{\ \alpha}^{a}e_{\ \beta}^{c}\frac{\delta\tilde{A}_{c}^{\prime}(q,\zeta)}{\delta\tilde{A}_{b}(q,\zeta)}\right|_{\zeta=0}\,,
\end{equation}
where we have neglected the inhomogeneity of the metric under the assumption of $\Omega\ll\omega,\,|{\bf q}|$ as discussed in Sec. \ref{sec:EOM}. The metric $g^{\alpha\beta}$, the vielbein $e^a_{\ \alpha}$, and vector fields are evaluated at a given coordinate $x$, $y$, relating to the position of the vector meson under consideration.
The numerator contains $\tilde{A}_a^\prime$, which can be expressed as linear combination of $E_i^\prime$. From the constraint equation (\ref{eq:constraint-E}), we derive that
\begin{equation}\label{eq:A-t-prime}
\tilde{A}_{t}^{\prime}(q,\zeta)=\frac{E_t^\prime(q,\zeta)}{\omega}=-\frac{\boldsymbol{\mathcal{D}}\cdot{\bf E}^{\prime}(q,\zeta)}{\omega\mathcal{D}_{t}}\,,
\end{equation}
where $\mathcal{D}_t$ and $\boldsymbol{\mathcal{D}}=(\mathcal{D}_x,\,\mathcal{D}_y,\,\mathcal{D}_z)$ are components of the row vector $\mathcal{D}$. On the other hand, according to the definition of electric fields in Eq. (\ref{eq:def-E}), we obtain 
\begin{equation}\label{eq:A-i-prime}
\tilde{A}_{i}^{\prime}(q,\zeta)=\frac{1}{\omega}E_{i}^{\prime}(q,\zeta)-\frac{q_{i}}{\omega}\tilde{A}_{t}^{\prime}(q,\zeta)\,,
\end{equation}
where $\tilde{A}_{t}^{\prime}(q,\zeta)$ on the right-hand-side is expressed in terms of $E_{i}^{\prime}(q,\zeta)$ as shown in Eq. (\ref{eq:A-t-prime}). Therefore the current correlation function in Eq. (\ref{eq:current-correlation}) can be evaluated via $\delta E_i^\prime(q,0)/\delta\tilde{A}_a(q,0)$.

Considering an arbitrary incoming wave solution of $E_i(q,\zeta)$, we can always find a set of basis $E_i^\text{nb}(q,e_j,\zeta)$ satisfying 
\begin{equation}\label{eq:nb-solution-behavior}
\lim_{\zeta\rightarrow 0}E_i^\text{nb}(q,e_j,\zeta)=\delta_{ij}\,,
\end{equation}
which allows us to express $E_i(q,\zeta)$ as a linear combination of the basis,
\begin{equation}\label{eq:general-solution}
E_i(q,\zeta)=\sum_j\,f_j\,E_i^\text{nb}(q,e_j,\zeta)\,.
\end{equation}
The coefficients $f_j=E_j(q,0)$ are determined by the values of the electric fields on the boundary. In the previous section in Eq. (\ref{eq:solution-near-horizon}), we have derived solutions $E_i^\text{nh}$ which have a series form near the horizon. The basis $E_i^\text{nb}$ can be written as a linear combination of $E_i^\text{nh}$ as follows,
\begin{equation}
E_i^\text{nb}(q,e_j,\zeta)=\sum_{s=0,\pm 1}c_{j,s}(q)E_i^\text{nh}(q,\zeta)\,,
\end{equation}
where the coefficients $c_{j,s}(q)$ are determined by matching the values at the boundary $\zeta=0$. We now consider a nonzero variation of $\tilde{A}_t(q,0)$. As a consequence, the variation of electric fields at the boundary reads,
\begin{equation}
\delta E_i(q,0)=q_i\delta\tilde{A}_t(q,0)
\end{equation}
With the help of Eq. (\ref{eq:general-solution}), we obtain the variation of $\delta E_i(q,\zeta)$,
\begin{equation}\label{eq:variation-Ei-At}
\delta E_{i}(q,\zeta)=E_{i}^\text{nb}(q,e_{j},\zeta)q_{j}\delta \tilde{A}_{t}(q,0)\,.
\end{equation}
Using Eqs. (\ref{eq:A-t-prime}), (\ref{eq:A-i-prime}), and (\ref{eq:variation-Ei-At}) we derive the following relations
\begin{align}\label{eq:dA-dA-1}
\frac{\delta\tilde{A}_{t}^{\prime}(q,0)}{\delta\tilde{A}_{t}(q,0)}=&-\frac{q_{j}\mathcal{D}_{i}}{\omega\mathcal{D}_{t}}\left.\partial_\zeta E_{i}^\text{nb}(q,e_{j},\zeta)\right|_{\zeta=0}\,, \nonumber\\
\frac{\delta\tilde{A}_{i}^{\prime}(q,0)}{\delta\tilde{A}_{t}(q,0)}=&\frac{q_{j}}{\omega}\left(\delta_{ik}+\frac{q_{i}\mathcal{D}_{k}}{\omega\mathcal{D}_{t}}\right)\left.\partial_\zeta E_{k}^\text{nb}(q,e_{j},\zeta)\right|_{\zeta=0}\,.
\end{align}
Similarly, a variation $\delta \tilde{A}_i(q,0)$ correspond to
\begin{equation}
\delta E_{i}(q,\zeta)=\omega E^\text{nb}_{i}(q,e_{j},\zeta)\delta \tilde{A}_j(q,0)\,,
\end{equation}
which leads to the following results
\begin{align}\label{eq:dA-dA-2}
\frac{\delta\tilde{A}_{t}^{\prime}(q,0)}{\delta\tilde{A}_{j}(q,0)}=&-\frac{\mathcal{D}_{i}}{\mathcal{D}_{t}}
\left.\partial_\zeta E^\text{nb}_{i}(q,e_{j},\zeta)\right|_{\zeta=0}\,, \nonumber\\
\frac{\delta\tilde{A}_{i}^{\prime}(q,0)}{\delta\tilde{A}_{j}(q,0)}=&\left(\delta_{ik}+\frac{q_{i}\mathcal{D}_{k}}{\omega\mathcal{D}_{t}}\right)\left.\partial_\zeta E^\text{nb}_{i}(q,e_{j},\zeta)\right|_{\zeta=0}\,.
\end{align}
Substituting Eqs. (\ref{eq:dA-dA-1}) and (\ref{eq:dA-dA-2}) into Eq. (\ref{eq:current-correlation}), we finally obtain the current correlation function $D^{ab}(x,y;q)$ for a vector meson with momentum $q_\mu$ and located at $(x,\,y)$ on the transverse plane. The spectral function is then given by the imaginary part of the current correlation function projected onto the polarization vectors
\begin{equation}\label{eq:spectral-functions}
\varrho_\lambda(x,y;q)=-\text{Im}\left[\epsilon_a^\ast(\lambda,q) \epsilon_b(\lambda,q)D^{ab}(x,y;q)\right]\,,
\end{equation}
where the polarization vector is 
\begin{equation}
\epsilon_a(\lambda,q)=\left(-\frac{{\bf q}\cdot\boldsymbol{\epsilon}_\lambda}{M},\boldsymbol{\epsilon}_\lambda+\frac{{\bf q}\cdot\boldsymbol{\epsilon}_\lambda}{M(\omega+M)}{\bf q}\right)\,,
\end{equation}
with $M\equiv\sqrt{\omega^2-{\bf q}^2}$ being the invariant mass. Here $\boldsymbol{\epsilon}_\lambda$ denotes the spin direction in the meson's rest frame. If the spin quantization direction is set to the direction of global rotation, i.e., the $z$-direction, the most natural choice for $\boldsymbol{\epsilon}_\lambda$ reads
\begin{align}
\boldsymbol{\epsilon}_0=&(0,0,1)\,,\nonumber\\
\boldsymbol{\epsilon}_1=&(-1,-i,0)/\sqrt{2}\,,\nonumber\\
\boldsymbol{\epsilon}_{-1}=&(1,-i,0)/\sqrt{2}\,.
\end{align}
In this case, the subscript $\lambda=S_z=0,\,\pm 1$ denotes the spin projection in the $z$-direction. We note that the result of the spectral function contains an ultraviolet divergence, which can be removed by a holographic renormalization. The presence of such divergence is because the solution near the boundary behaves as 
\begin{align}
E^\text{nb}_i(q,e_j,\zeta)\approx& \delta_{ij} +C_{ij}^{(1)} \left[\zeta^2 +\mathcal{O}(\zeta^4)\right]\nonumber\\ 
&+ C_{ij}^{(2)}\left[\zeta^2+\mathcal{O}(\zeta^4)\right]\ln\zeta \,,
\end{align}
where the term of $C_{ij}^{(2)}$ arises from the non-normalizable mode. Numerically, we remove this ultraviolet divergence by fitting $E^\text{nb}_i(q,e_j,\zeta)$ with $\delta_{ij} +C_{ij}^{(1)}\zeta^2+C_{ij}^{(2)}\zeta^2\ln\zeta$ in a small region around $\zeta=0$ and then setting $C_{ij}^{(2)}=0$.

\section{Numerical results}\label{sec:Numerical-results}

\subsection{Momentum parallel to rotating axis}\label{sec:parallel-case}
In the absence of global rotation, the current correlation function can be decomposed into contributions from spectral functions for transversely polarized and longitudinally polarized states, 
\begin{align}
D^{ab}(q)=&-\rho_L(q)\epsilon^a_h(q)\epsilon^b_h(q)\nonumber\\
&+\rho_T(q)\left[g^{ab}-\frac{q^a q^b}{q^2}+\epsilon^a_h(q)\epsilon^b_h(q)\right]\,,
\end{align}
where $\epsilon^a_h(q)$ is the polarization vector for the spin-0 state in the helicity frame, 
\begin{equation}\label{eq:epsilon_h}
\epsilon^a_h(q)=\left(-\frac{|{\bf q}|}{M},\,\frac{\omega\,{\bf q}}{M|{\bf q}|}\right)\,.
\end{equation}
When the transverse momentum of the meson vanishes, i.e., ${\bf q}=(0,0,q_z)$, the polarization vector (\ref{eq:epsilon_h}) coincides with that of the state with spin projection $S_z=0$. This implies that the longitudinally polarized state is also an eigenstate of spin with $(S,S_z)=(1,0)$. The transversely polarized states, on the other hand, are eigenstates of spin with $(S,S_z)=(1,\pm 1)$. In the absence of rotation, $\Omega=0$, the $(1,\pm 1)$ states are degenerate and share the same spectral function $\rho_T(q)$, while the $(1,0)$ state has spectral function $\rho_L(q)$. In general, $\rho_T\neq\rho_L$ if $q_z\neq 0$ because the motion breaks the symmetry between transverse and longitudinal directions. 

When the global rotation is present, the energy of the meson will be shifted by the coupling between the rotation and the total angular momentum. A non-relativistic prediction gives \citep{mashhoon1988neutron,mashhoon1998observable,hehl1990inertial}
\begin{equation}\label{eq:classical-relation}
\omega(\Omega)=\omega(0)-(L_z+S_z)\Omega\,,
\end{equation}
where $\omega(\Omega)=\sqrt{[M(\Omega)]^2+{\bf q}^2}$ denotes the energy as a function of the angular velocity $\Omega$, and $L_z,\,S_z$ denote the orbital and spin angular momenta with respect to the rotation axis. For a meson moving along the $z$-direction, the orbital part vanishes, $L_z=0$, and thus the energy shift arises solely from the spin-rotation coupling.

\begin{figure}[tbh]
\includegraphics[width=0.8\linewidth]{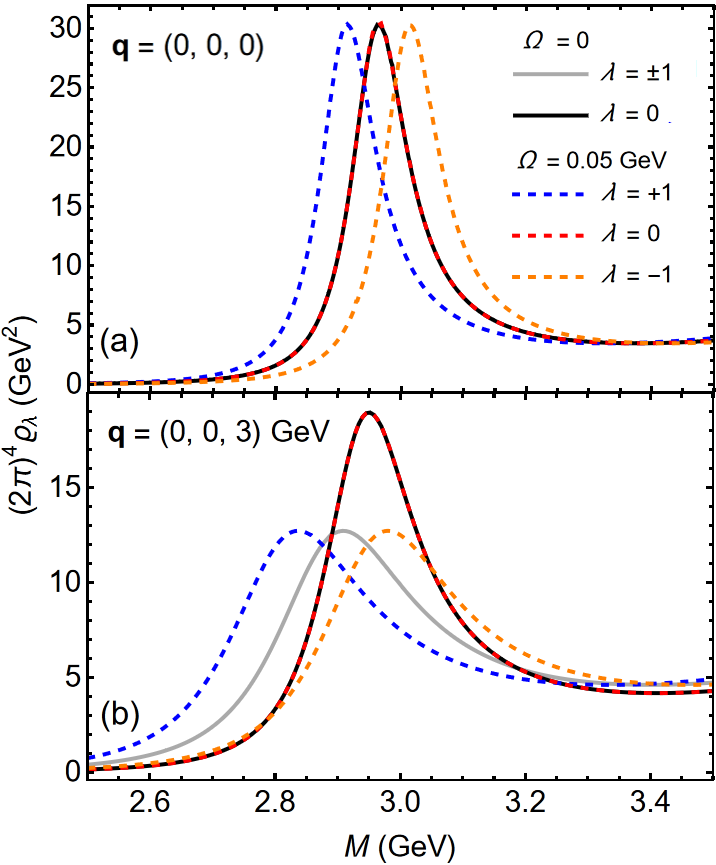}\\
\caption{Invariant-mass spectral functions for $J/\psi$ at $T=150$ MeV. The particle's momentum is set to ${\bf q}=(0,0,0)$ [panel (a)] or ${\bf q}=(0,0,3)$ GeV [panel (b)], while the global angular velocity $\Omega=0$ (solid lines) or $\Omega=0.05$ GeV (dashed lines).}\label{fig:spectral-function}
\end{figure}

\begin{figure}[tbh]
\includegraphics[width=0.8\linewidth]{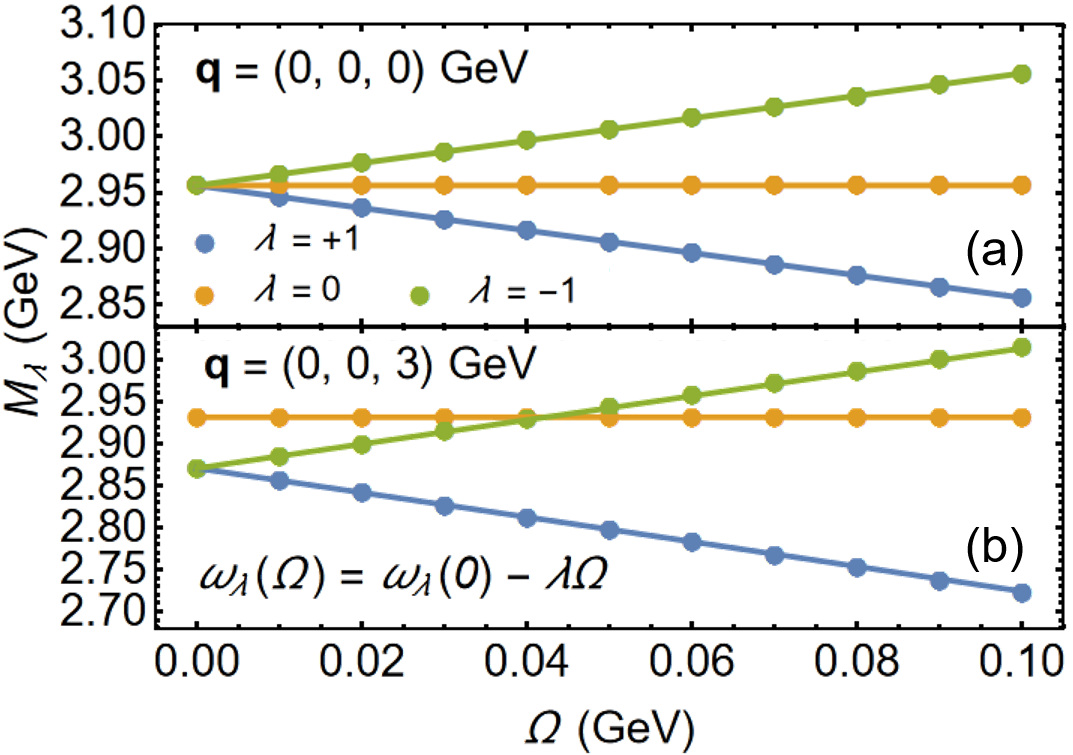}
\caption{Masses of $J/\psi$ at $T=150$ MeV as functions of $\Omega$. Results for $\lambda=+1,\,0,\,-1$ states are shown by blue, orange, and green colors, respectively, with dots representing results of the holographic model and solid lines being determined by Eq. (\ref{eq:classical-relation}).}\label{fig:Mass-Omega-1}
\end{figure}

\begin{figure}[tbh]
\includegraphics[width=0.8\linewidth]{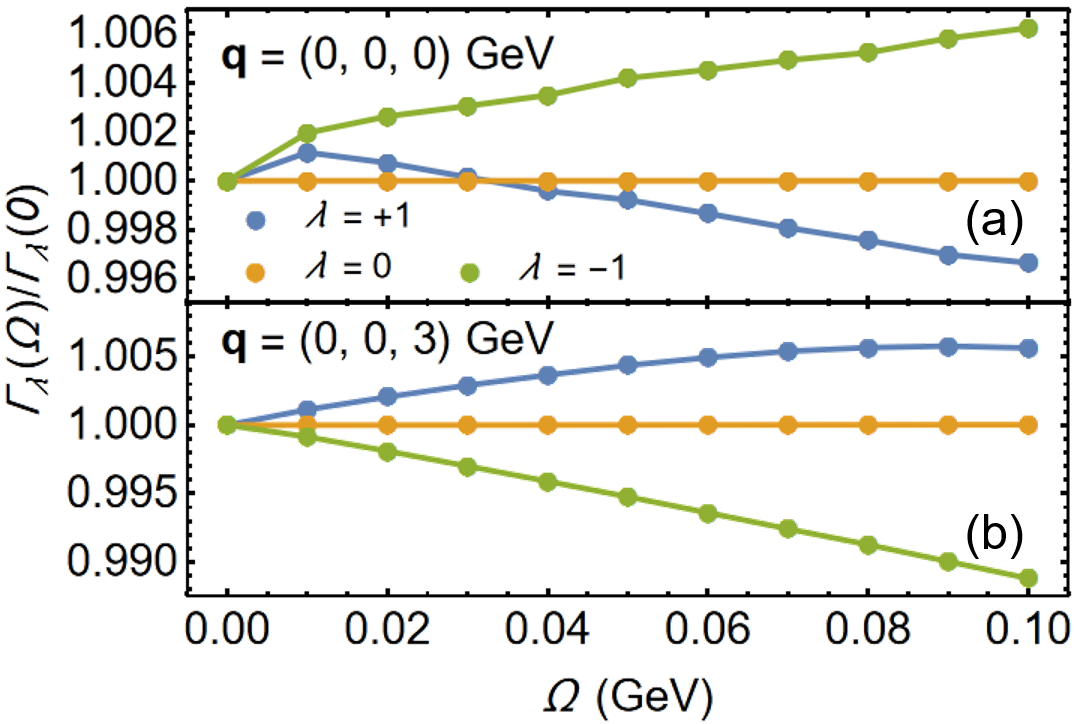}
\caption{Width ratios $\Gamma_\lambda(\Omega)/\Gamma_\lambda(0)$ of $J/\psi$ at $T=150$ MeV as functions of $\Omega$. Results for $\lambda=+1,\,0,\,-1$ states are shown by blue, orange, and green colors, respectively.}\label{fig:Gamma-Omega-1}
\end{figure}

Within the holographic framework, we numerically calculate the invariant-mass spectral functions of the $J/\psi$ meson at $T=150$ MeV, as shown in Fig. \ref{fig:spectral-function}. When both the global rotation and the meson's momentum vanish, the spectral functions are three-fold degenerate, as indicated by the black line in Fig. \ref{fig:spectral-function} (a). A nonzero $\Omega$ leads to splittings between three spin states: the $\lambda=+1$ ($\lambda=-1$) state shifts to lower (higher) mass while the $\lambda=0$ state remains unchanged. This behavior is quantitatively consistent with the relation (\ref{eq:classical-relation}). The spectral functions in Fig. \ref{fig:spectral-function} are independent to the transverse location $(x,y)$.

In order to check the relation (\ref{eq:classical-relation}) quantitatively, we fit the spectral function in the interval $2.5 \text{ GeV}<M<3.5$ GeV using \citep{Fujita:2009wc}
\begin{equation}\label{eq:fit-function}
\rho_\lambda(M)=\frac{a(M/\text{GeV})^b}{(M-M_\lambda)^2+\Gamma_\lambda^2}\text{GeV}^4,
\end{equation}
where $a$ and $b$ are dimensionless parameters, $M_\lambda$ describes the mass of spin-$\lambda$ state and $\Gamma_\lambda$ its width. The extracted $M_\lambda$ as functions of $\Omega$ are shown in Fig. \ref{fig:Mass-Omega-1}, where the points represent the numerical results from the holographic framework and the solid lines are given by Eq. (\ref{eq:classical-relation}). The excellent agreement between the points and solid lines demonstrates that the prediction from the rotation-angular momentum coupling is well reproduced in the holographic model. On the other hand, Fig. \ref{fig:Gamma-Omega-1} shows the width $\Gamma_\lambda(\Omega)$ normalized by its value at $\Omega=0$. We find that the widths depend very weakly on $\Omega$. Up to $\Omega=0.1$ GeV, the variation is less than $1\%$. We therefore conclude that, when the meson's momentum is parallel to the rotation axis, the global rotation shifts the spectral functions without appreciably changing their shape.

\subsection{Momentum perpendicular to rotating axis}

\begin{figure}[tbh]
\includegraphics[width=0.8\linewidth]{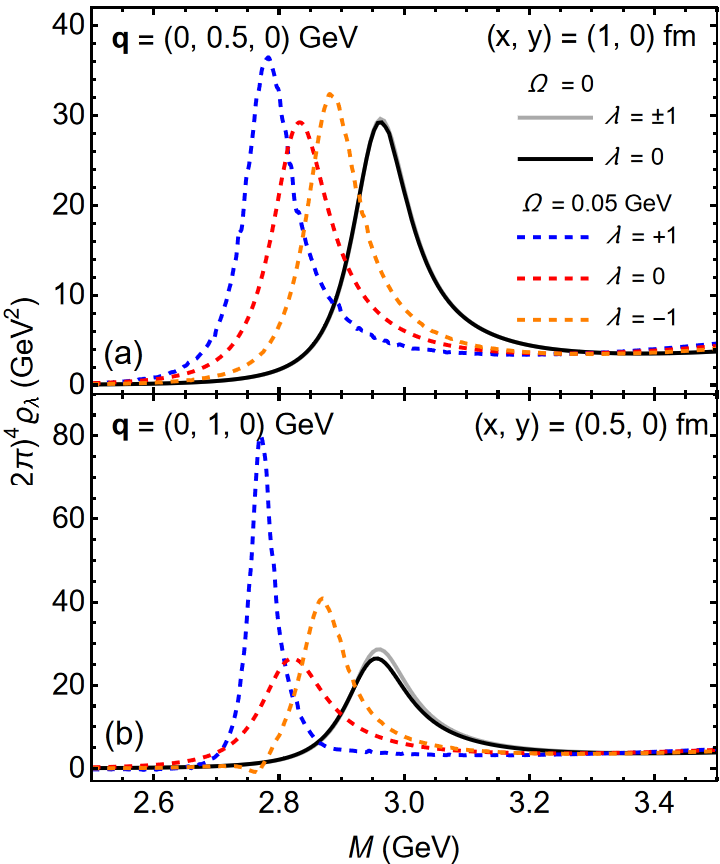}
\caption{Invariant-mass spectral functions for $J/\psi$ with momentum ${\bf q}=(0,0.5,0)$ GeV [panel (a)] or ${\bf q}=(0,1,0)$ GeV [panel (b)], while the transverse locations are set to $(x,y)=(1,0)$ fm [panel (a)] and $(0.5,0)$ fm [panel (b)], respectively. For these two cases, the orbital angular momenta $L_z=q_yx$ are the same. Results in the absence of rotation and $\Omega=0.05$ GeV are shown by solid lines and dashed lines, respectively.} \label{fig:small-qt}
\end{figure}

We now consider the case when the meson's momentum is perpendicular to the rotation axis. Without loss of generality, we set $q_x=0$ and $q_y\neq 0$. For a meson located at $(x,y)$ on the transverse plane, the orbital angular momentum is $L_z=x q_y$, leading to a sizable contribution to the rotational energy shift in Eq. (\ref{eq:classical-relation}). Figure \ref{fig:small-qt} (a) shows the invariant-mass from the peak of the spectral functions  for mesons with $q_y=0.5$ GeV located at $(x,y)=(1,0)$ fm. Similar to the case discussed in Sec. \ref{sec:parallel-case}, the global rotation breaks the degeneracy among the three spin states. Compared to the results at $\Omega=0$ [solid lines in Fig. \ref{fig:small-qt} (a)], those at $\Omega=0.05$ GeV (dashed lines) shift toward lower mass. A similar pattern is observed for mesons with $q_y=1$ GeV located at $(x,y)=(0.5,0)$ fm, as shown in Fig. \ref{fig:small-qt} (b). In both of the above two cases, the orbital angular momentum is the same $L_z=0.5\,\text{GeV/c}\cdot\text{fm}=2.538 \hbar$. 

\begin{figure}[tbh]
\includegraphics[width=0.8\linewidth]{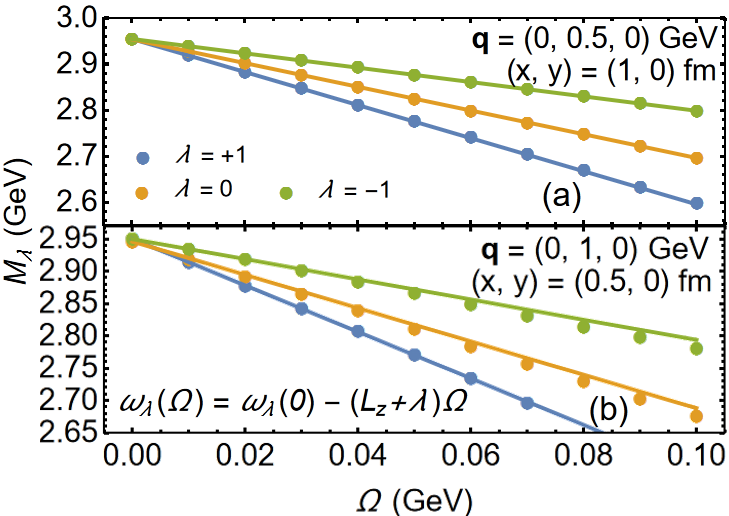}
\caption{Masses of $J/\psi$ with transverse momentum $q_y=0.5$ GeV [panel (a)] or 1 GeV [panel (b)] as functions of $\Omega$.} \label{fig:small-qt-Mass-Omega}
\end{figure}

\begin{figure}[tbh]
\includegraphics[width=0.8\linewidth]{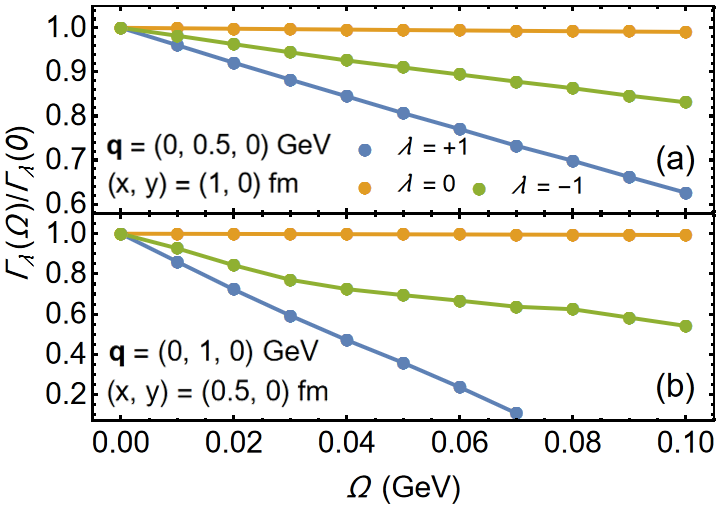}
\caption{Width ratios $\Gamma_\lambda(\Omega)/\Gamma_\lambda(0)$ of $J/\psi$ with transverse momentum $q_y=0.5$ GeV [panel (a)] or 1 GeV [panel (b)] as functions of $\Omega$.} \label{fig:small-qt-Gamma-Omega}
\end{figure}

The masses and widths are extracted by fitting the spectral functions with Eq. (\ref{eq:fit-function}), and the results are shown in Figs. \ref{fig:small-qt-Mass-Omega} and \ref{fig:small-qt-Gamma-Omega}, respectively. We find a mass ordering $M_{-1}>M_0>M_{+1}$ due to the rotational energy shift. Holographic results of masses for mesons with $q_y=0.5$ GeV [dots in Fig. \ref{fig:small-qt-Mass-Omega} (a)] agree very well with the values predicted by Eq. (\ref{eq:classical-relation}) [solid lines in Fig. \ref{fig:small-qt-Mass-Omega} (a)]. This is mainly because the momentum $q_y=0.5$ GeV is much smaller than the mass and the meson remains close to a non-relativistic particle. However, when $q_y=1$ GeV, as shown in Fig. \ref{fig:small-qt-Mass-Omega} (b), the meson masses begin to deviate from Eq. (\ref{eq:classical-relation}) at large angular velocity, e.g., $\Omega>0.05$ GeV. The meson widths are shown in Fig. 
\ref{fig:small-qt-Gamma-Omega} as functions of $\Omega$. As $\Omega$ increases, the widths of the $\lambda=\pm 1$ states decrease significantly, while the width of the $\lambda=0$ state remains nearly unchanged. 

\begin{figure}[tbh]
\includegraphics[width=0.8\linewidth]{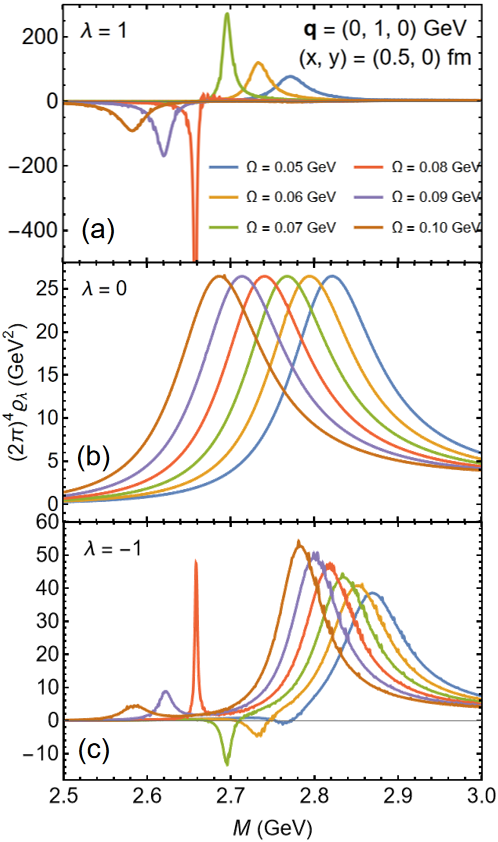}
\caption{Invariant-mass spectral functions for $J/\psi$ with momentum ${\bf q}=(0,1,0)$ GeV at $\Omega=0.05$-0.1 GeV. Spin $\lambda=1,0,-1$ states are shown in panels (a), (b), and (c), respectively. Small oscillations originate from numerical noise.} \label{fig:large-qt-1}
\end{figure}

We note that when $q_y=1$ GeV and $\Omega>0.07$ GeV, the spectral functions of the $\lambda= 1$ state deviate from the single-peak form, so it cannot be fitted using Eq. (\ref{eq:fit-function}). For this reason, the corresponding results are not shown in Figs. \ref{fig:small-qt-Mass-Omega} (b) and \ref{fig:small-qt-Gamma-Omega} (b). To illustrate the change in spectral shape more clearly, we show in Fig. \ref{fig:large-qt-1} the spectral functions of different spin states at $\Omega=0.05$-0.1 GeV. For the $\lambda=1$ state, the spectral function $\varrho_1$ exhibits a single-peak whose position moves toward lower mass as $\Omega$ increases, as predicted by Eq. (\ref{eq:classical-relation}). When $\Omega\geq 0.08$ GeV, $\varrho_1$ flips its sign and becomes negative. For the $\lambda=-1$ state, two peaks appear when $\Omega\geq 0.06$ GeV: the peak at larger mass follows the relation (\ref{eq:classical-relation}), while the other peak is similar to $\varrho_{1}$ but with opposite sign. 

\begin{figure}[tbh]
\includegraphics[width=0.8\linewidth]{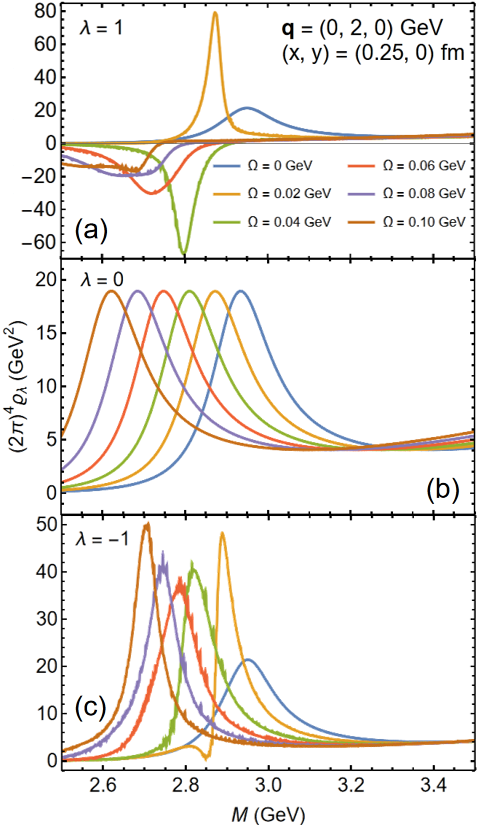}
\caption{Invariant-mass spectral functions for $J/\psi$ with momentum ${\bf q}=(0,2,0)$ GeV at $\Omega=0$-0.1 GeV. Spin $\lambda=1,0,-1$ states are shown in panels (a), (b), and (c), respectively. Small oscillations originate from numerical noise.} \label{fig:large-qt-2}
\end{figure}

The negative spectral functions and multi-peak structures at large $q_y$ and large $\Omega$ indicate that the rotational effect is no longer a simple energy shift induced by the coupling $-\Omega J_z$. Instead, the projected retarded Green function, i.e., $\epsilon_a^\ast(\lambda,q) \epsilon_b(\lambda,q)D^{ab}(x,y;q)$ in Eq. (\ref{eq:spectral-functions}) is not diagonal. At the operator level, the total angular momentum $J_z=L_z+S_z$ does not commute with the transverse momentum, indicating that eigenstates cannot have definite $J_z$ and ${p_x,\,p_y}$ at the same time. In other words, fixed-momentum states are not eigenstates of $J_z$. As a consequence, projecting the fixed-momentum states onto baiss with fixed-$J_z$ leads to off-diagonal components of the projected retarded Green function. The negative values in the spectral function should therefore not be interpreted as negative physical densities of propagating modes, but rather substantial distortions due to the interference effect. This effect becomes stronger for larger transverse momentum. As shown in Fig. \ref{fig:large-qt-2}, for mesons with $q_y=2$ GeV, $\varrho_1$ becomes negative already at $\Omega\geq 0.04$ GeV, which is smaller than the corresponding value ($\Omega\geq0.08$ GeV)
in Fig. \ref{fig:large-qt-1}. By contrast, in both Fig. \ref{fig:large-qt-1} and \ref{fig:large-qt-2}, the spectral function for the $\lambda=0$ state always shifts with $\Omega$ while keeping its shape unchanged because the $\lambda=0$ state does not mix with other states. 

\section{Summary}\label{sec:summary}

In this work, we investigate the spectral properties of $J/\psi$ mesons in a rotating thermal background within the soft-wall holographic QCD model.
The global rotation is included in the curved metric, while the local inertial frame is introduced in which the meson can have well-defined spins.
Equations of motion for the vector field in the local inertial frame are derived and solved analytically near the horizon up to the next-to-leading order in $\zeta/\zeta_h-1$. The retarded Green function for the meson, according to the gauge/gravity correspondence, is calculated by values of incoming-wave solutions and their derivatives with respect to the bulk coordinate at the boundary. These values are evaluated numerically by solving the equations of motion together with solutions near the horizon as boundary conditions. Spectral functions of the $J/\psi$ meson are extracted by projecting the retarded Green function onto polarization vectors with definite spins and taking the imaginary part. 

Taking $J/\psi$ at $T=150$ MeV as an example, we show the spectral functions for different momenta, angular velocities, and spins. The results demonstrate that the global rotation does have a significant impact on the in-medium spectral functions. The peak positions exhibit an energy shift arising from the coupling between rotation and the meson's total angular momentum. If the momentum is parallel to the rotation axis, the energy shift is linear in the angular velocity, i.e., $-\Omega J_z$, which coincides with the non-relativistic prediction, and the width of the spectral function is nearly independent of $\Omega$. On the other hand, if the momentum is perpendicular to the rotation axis, different spin states exhibit different behaviors and the rotational effects become more prominent for the $\lambda=\pm 1$ states. Spectral functions could change their shapes and even become negative when $\Omega$ and the transverse momentum are large enough. Such behaviors are due to the interference between the $\lambda=\pm 1$ states and may offer theoretical insights for studies on the spin alignment of vector meson in a rotating system. 

Compared with previous works \cite{Braga:2023fac,Zhao:2023pne,Zhu:2024uwu,Wang:2024rim}, the present work provides a systematic investigation of $J/\psi$ with definite spin by introducing the local inertial frame. This framework successfully reproduces the clear triplet splitting of spectral functions which are not shown in other holographic studies. The whole framework can be applied to studying $\phi$ and bottomonium properties by choosing an appropriate parameter in the dilaton field. Future studies based the present framework, such as the effects of baryon chemical potential, strong magnetic field, or anisotropy of the background, may provide deeper insight into the role of vorticity in heavy quarkonium properties in the quark-gluon plasma.

\begin{acknowledgments}
The authors thank Si-Wen Li and Yan-Qing Zhao for helpful discussions. X.L.S is supported by the National Natural Science Foundation of China under Grant No. 12547102 and No. 12147101. D.H. is supported in part by the National Natural Science Foundation of China(NSFC) under Grants No. 12435009, and No. 12275104. H.C.R. is supported in part by the National Key Research and Development Program of China
under Contract No. 2022YFA1604900. The data that support the findings of this article are
openly available \citep{sheng_2026_21739915}.
\end{acknowledgments}

\bibliographystyle{apsrev4-1}
\bibliography{bibfile}

\onecolumngrid
\appendix

\section{Explicit expressions}\label{Appendix}

In this appendix, we collect explicit expressions of coefficient matrices appearing in Sec. \ref{sec:Near-Horizon}. The matrices $\mathcal{T}_{1,2,3}$ in Eq. \eqref{eq:EOM-A} are given by
\begin{align}
\mathcal{T}_{1}=&\frac{\zeta^{4}Q(\zeta)}{L^{4}}\left(\begin{array}{cccc}
-1 & 0 & 0 & 0\\
-\Omega y & f(\zeta) & 0 & 0\\
\Omega x &  & f(\zeta) &0\\
0 & 0 & 0 & f(\zeta) 
\end{array}\right)\,, \nonumber \\
\mathcal{T}_{2}=&\frac{\zeta^{4}Q(\zeta)}{L^{4}}\left(-\frac{1}{\zeta}-2c\zeta\right)\left(\begin{array}{cccc}
-1 & 0 & 0 & 0\\
-\Omega y & 1+\frac{(3-2c\zeta^{2})\zeta^{4}}{(1+2c\zeta^{2})\zeta_{h}^{4}} & 0 &0\\
\Omega x & 0 & 1+\frac{(3-2c\zeta^{2})\zeta^{4}}{(1+2c\zeta^{2})\zeta_{h}^{4}} & 0\\
0 & 0 & 0 & 1+\frac{(3-2c\zeta^{2})\zeta^{4}}{(1+2c\zeta^{2})\zeta_{h}^{4}}
\end{array}\right)\,, \nonumber\\
\mathcal{T}_{3}=&\frac{\zeta^{4}Q(\zeta)}{L^{4}f(\zeta)}\left(\begin{array}{cccc}
{\bf q}^{2} & \omega q_{x} & \omega q_{y} & \omega q_{z}\\
\omega q_{x} & \omega^{2}-{\bf q}^{2}f(\zeta)+q_{x}^{2}f(\zeta) & q_{x}q_{y}f(\zeta) & q_{x}q_{z}f(\zeta)\\
\omega q_{y} & q_{x}q_{y}f(\zeta) & \omega^{2}-{\bf q}^{2}f(\zeta)+q_{y}^{2}f(\zeta) & q_{y}q_{z}f(\zeta)\\
\omega q_{z} & q_{x}q_{z}f(\zeta) & q_{y}q_{z}f(\zeta) & \omega^{2}-{\bf q}^{2}f(\zeta)+q_{z}^{2}f(\zeta)
\end{array}\right)\nonumber \\
&+\frac{\zeta^{4}Q(\zeta)}{L^{4}f(\zeta)}\Omega\left(\begin{array}{cccc}
0 & q_{x}L_{z} & q_{y}L_{z} & q_{z}L_{z}\\
-iq_{y}+y{\bf q}^{2}+q_{x}L_{z} & 2\omega L_{z}+y\omega q_{x} & -2i\omega+y\omega q_{y} & y\omega q_{z}\\
iq_{x}-x{\bf q}^{2}+q_{y}L_{z} & 2i\omega-x\omega q_{x} & 2\omega L_{z}-x\omega q_{y} & -x\omega q_{z}\\
q_{z}L_{z} & 0 & 0 & 2\omega L_{z}
\end{array}\right) \nonumber \\
&+\frac{\zeta^{4}Q(\zeta)}{L^{4}f(\zeta)}\Omega^{2}\left(\begin{array}{cccc}
0 & 0 & 0 & 0\\
0 & 1+L_{z}^{2}+i(xq_{x}+yq_{y})+yq_{x}L_{z} & -2iL_{z}+yq_{y}L_{z} & yq_{z}L_{z}\\
0 & 2iL_{z}-xq_{x}L_{z} & 1+L_{z}^{2}+i(xq_{x}+yq_{y})-xq_{y}L_{z} & -xq_{z}L_{z}\\
0 & 0 & 0 & L_{z}^{2}+i(xq_{x}+yq_{y})
\end{array}\right)\,,
\end{align}
where $L_z=xq_y-yq_x$ is the orbital angular momentum in the direction of global rotation. In Eq. \eqref{eq:constraint-psi}, the coefficient $\mathcal{D}$ haas the following form
\begin{equation}
\mathcal{D}=\frac{\zeta^4 Q(\zeta)f(\zeta)}{\omega L^4}\left(\frac{\omega(\omega+\Omega L_z)-{\bf q}^2f(\zeta)}{\omega f(\zeta)},\,q_x,\,q_y,\,q_z\right)\,.
\end{equation}
Near the horizon we expand $\mathcal{D}$ into a Taylor expansion as shown in Eq. \eqref{eq:expansion-F-G-D}. The leading order term, $\mathcal{D}^{(0)}$, is given in Eq. \eqref{eq:coefficients-G2-D0}, while the next-to-leading order term reads
\begin{equation}
\mathcal{D}^{(1)}=-\frac{4\zeta_h^4 Q(\zeta_h)}{\omega L^4} \left(\frac{\omega^2-4{\bf q}^2+\omega\Omega L_z+2c\zeta_h^2\omega(\omega+\Omega L_z)}{4\omega},\,q_x,\,q_y,\,q_z\right)\,.
\end{equation}
When solving the vector field near the horizon, we also need to know the coefficients $\tilde{\mathcal{F}}^{-1}$,  $\tilde{\mathcal{G}}^{-2}$, and  $\tilde{\mathcal{G}}^{-1}$. The explicit expression for $\tilde{\mathcal{G}}^{-2}$ is given in Eq. \eqref{eq:expansion-F-G-D}. The remaining two coefficients are
\begin{align}
\tilde{\mathcal{F}}^{(-1)}=&\frac{1}{\zeta_{h}}\left(\begin{array}{cccc}
-\frac{i\tilde{\omega}\zeta_{h}}{2} & 0 & 0 & 0\\
-\frac{q_{x}}{\omega} & 1-\frac{i\tilde{\omega}\zeta_{h}}{2} & 0 & 0\\
-\frac{q_{y}}{\omega} & 0 & 1-\frac{i\tilde{\omega}\zeta_{h}}{2} & 0\\
-\frac{q_{z}}{\omega} & 0 & 0 & 1-\frac{i\tilde{\omega}\zeta_{h}}{2}
\end{array}\right)+\mathcal{O}(\Omega^2)\,,\\
\tilde{\mathcal{G}}^{(-1)}=&
\frac{1}{16\zeta_{h}}\left(\begin{array}{cccc}
\tilde{\omega}(3\tilde{\omega}\zeta_{h}+8ic\zeta_{h}^{2}+4i) & 4\omega q_{x}\zeta_{h} & 4\omega q_{y}\zeta_{h} & 4\omega q_{z}\zeta_{h}\\
0 & C & 0 & 0\\
0 & 0 & C & 0\\
0 & 0 & 0 & C
\end{array}\right)\nonumber\\
&+\frac{\Omega L_{z}}{4\omega}\left(\begin{array}{cccc}
\omega^{2} & \omega q_{x} & \omega q_{y} & \omega q_{z}\\
\omega q_{x} & q_{x}^{2} & q_{x}q_{y} & q_{x}q_{z}\\
\omega q_{y} & 4q_{x}q_{y} & q_{y}^{2} & q_{y}q_{z}\\
\omega q_{z} & 4q_{x}q_{z} & q_{y}q_{z} & q_{z}^{2}
\end{array}\right)-\frac{\Omega L_{z}}{16\omega^{2}}\left(\begin{array}{cccc}
4\omega(\omega^{2}+{\bf q}^{2}) & 0 & 0 & 0\\
(\omega^{2}+4{\bf q}^{2})q_{x} & 6\omega^{3} & 0 & 0\\
(\omega^{2}+4{\bf q}^{2})q_{y} & 0 & 6\omega^{3} & 0\\
(\omega^{2}+4{\bf q}^{2})q_{z} & 0 & 0 & 6\omega^{3}
\end{array}\right)\nonumber\\
&+\frac{3i}{16}\Omega\left(\begin{array}{cccc}
0 & 0 & 0 & 0\\
-q_{y} & 0 & 2\omega & 0\\
q_{x} & -2\omega & 0 & 0\\
0 & 0 & 0 & 0
\end{array}\right)+\mathcal{O}(\Omega^2)\,,
\end{align}
where
\begin{equation}
C\equiv4i\tilde{\omega}(1+2c\zeta_{h}^{2})+4{\bf q}^{2}\zeta_{h}-3\zeta_{h}(\omega^{2}-\tilde{\omega}^{2})
\end{equation}
and we have truncated $\mathcal{O}(\Omega^2)$ contributions.  

\end{document}